# The structure is the message: preserving experimental context through tensor decomposition

Zhixin Cyrillus Tan [a†], Aaron S. Meyer [a-d†]

[a] Bioinformatics Interdepartmental Program, University of California, Los Angeles (UCLA), USA
[b] Department of Bioengineering, UCLA, USA
[c] Jonsson Comprehensive Cancer Center, UCLA, USA
[d] Eli and Edythe Broad Center of Regenerative Medicine and Stem Cell Research, UCLA, USA
[†] Email: cyztan@gmail.com, ameyer@asmlab.org

## Abstract

Recent biological studies have been revolutionized in scale and granularity by multiplex and high-throughput assays. Profiling cell responses across several experimental parameters, such as perturbations, time, and genetic contexts, leads to richer and more generalizable findings. However, these multidimensional datasets necessitate a reevaluation of the conventional methods for their representation and analysis. Traditionally, experimental parameters are merged to flatten the data into a two-dimensional matrix, sacrificing crucial experiment context reflected by the structure. As Marshall McLuhan famously stated, “The medium is the message.” In this work, we propose that the experiment structure is the medium in which subsequent analysis is performed, and the optimal choice of data representation must reflect the experiment structure. We introduce tensor-structured analyses and decompositions to preserve this information. We contend that tensor methods are poised to become integral to the biomedical data sciences toolkit.

Multiplex and high-throughput assays now enable the exploration of cell responses in unprecedented scale and detail. Consequently, studies of biological systems have increasingly focused on profiling biological systems across multiple contexts. For instance, a panel of candidate therapies might be profiled using fibroblasts derived from multiple organs, with several features of their response measured over time (Fig. 1a). Identifying how responses are shared or distinct across multiple cellular contexts and experimental conditions reveals more about the biological mechanism and enhances the generalizability of the results. At the same time, measuring cells and tissues across multiple parameters generates data with multiple dimensions (e.g. cell line, time, experimental conditions), which necessitates reevaluating how we represent and analyze such information.

Representing multivariate data in a tabular form can sacrifice the ultimate insight that can be derived. It is not uncommon that studies with several dimensions are still laid out in rows and columns with some dimensions merged. For the example in Fig. 1a, when the experiment is repeated over time, the columns must expand to combine two experimental parameters, drug and time point, such as “alfazumab – 1 hr,” “alfazumab – 3 hr,” “bravociclib – 1 hr,” “bravociclib – 3 hr”, etc. In this format, one may instinctively apply familiar off-the-shelf statistical approaches, such as principal component analysis (PCA), because the data appears to be in matrix form.

So, what is the problem with this? As communication philosopher Marshall McLuhan famously stated[1], “The medium is the message.” The choice of data structure can bias its analysis. A tabular form implicitly treats each column and row as independent of one another, while merged dimensions diverge from this assumption. For instance, “alfazumab – 1 hr” and “alfazumab – 3 hr” share the same treatment, and “alfazumab – 1 hr” and “bravociclib – 1 hr” share the same timing; however, “bravociclib – 1 hr” and “alfazumab – 3 hr” differ in two distinct ways (Fig. 1b). When flattening a multidimensional dataset into a table, we compromise this property.

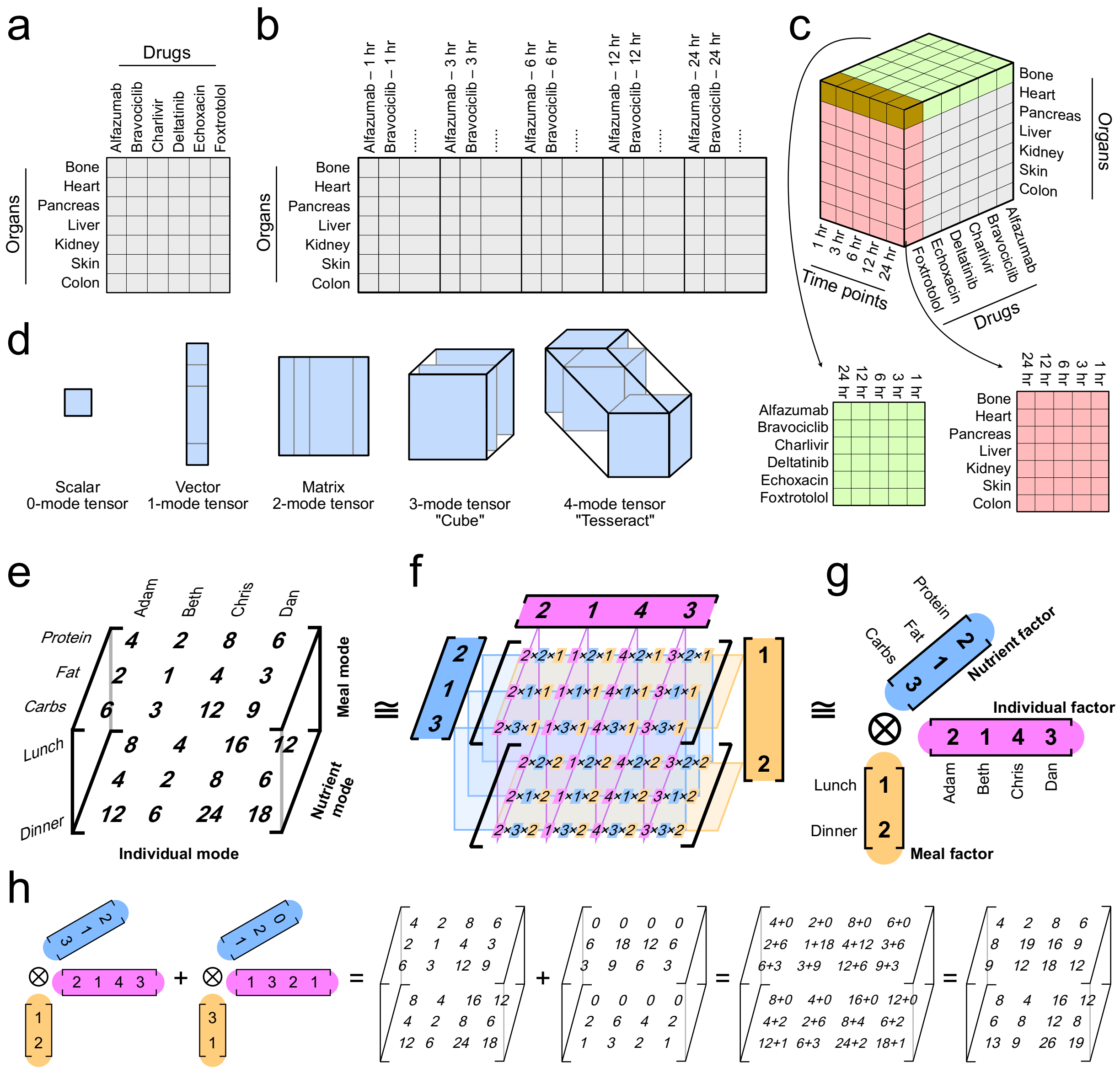

**Figure 1. Basic concepts of tensor-structured data**
a) A dataset on cells collected from different organs responding to various drug treatments can be documented by a table. b) When the measurements are performed over multiple time points, the original columns in the table can be expanded into multilevel indices, recording both drug and time. This nonetheless breaks the assumption that all columns are equally related. c) Alternatively, the same data can be recorded as a three-dimensional array, with organ, drug, and time point as three separate degrees of freedom. Here, the pink represents how every cell responds to foxtrotolol over time, and the green represents the pharmacokinetic profile of bone cells over all treatments. The brown is shared by pink and green. d) Tensors are multidimensional arrays. A dimension of a tensor is a mode. 0, 1, and 2-mode arrays are known as scalar, vector, and matrix. e) An example of a rank-one tensor. It describes the diets of four people and has dimensions 4×3×2, individual by nutrient by meal. f) Rank-one tensors are those whose every entry can be written as the product of a few numbers, one from each mode-specific vector, from their corresponding coordinates. g) A rank-one tensor can be written as multiple mode-specific factors joined by the vector outer product, ⊗. h) Even written as collections of vectors, these rank-one tensors should still be understood as arrays with numbers in every entry. Adding two tensors of the same shape is to add their corresponding position together.

To devise a more effective approach, the "medium", or structure, of the experiment must be incorporated. The example experiment varies across three degrees of freedom: organ, drug, and time point; this is best represented by a three-dimensional array or tensor (Fig. 1c). A tensor representation aligns entries with shared meaning. For instance, when examined from the perspective of an organ (e.g., bone, the green cubes), we find the pharmacokinetics profiles of all drugs on this organ; when viewed from a drug (e.g., foxtrotolol, the pink cubes), we find its impact on all organs over time (Fig. 1c).

In this work, we aim to introduce tensor methods to the wider systems biology community. We propose that tensor methods should and will become an established part of the basic biomedical data sciences toolbox.

## Defining tensors and tensor decomposition

Tensors are nothing more than multidimensional arrays[2–4]. Zero-, one- and two-dimensional tensors are scalars, vectors, and matrices, respectively (Fig. 1d). To avoid conflicting definitions of "dimension" in linear algebra, "mode" or "order" are used—three-dimensional, three-mode, and third-order tensors are all the same concepts. A matrix has two modes—columns and rows—but tensors over three modes can run out of words to describe their modes. When structuring biological data into a tensor, each mode ideally relates to a varied parameter of the experiment, such as samples, genes, cell lines, treatments, concentrations, or time points.

Tensors can be analyzed by tensor decomposition. Before describing this, it is helpful to introduce the concept of rank-one tensors, the building block for tensor decomposition. Like with matrices, even large data tensors can be decomposed into a series of simple patterns, known as rank-one tensors. Unlike the concept of tensor mode which is directly associated with the data dimensionality, the rank of a tensor is a separate and less evident concept that requires examining its entries. As an illustrative example, consider a dataset with the diets of four individuals, tracking the amount of three nutrients in two meals. By stacking the lunch nutrient facts (a 4×3 matrix) on top of dinner (another 4×3 matrix), we obtain a 4×3×2 tensor with individual, nutrient, and meal modes (Fig. 1e). In this contrived example, along the nutrient mode, every vector is a multiple of [2, 1, 3]. This indicates that all eight meals have the same nutritional composition. Dinner factor values are double those of lunches, suggesting that all dinners are twice as large. The individual factor is [2, 1, 4, 3], indicating the ratio of the four people's appetite: Adam eats twice as much as Beth eats in every meal, while Chris and Dan eat four times and three times as Beth eats, respectively. Every entry in this tensor can be precisely computed by multiplying three numbers, each from the nutrient, meal, and individual factor with their positions corresponding to its position in the tensor (Fig. 1f). To describe this property, we define this tensor as the outer product of these three vectors (Fig. 1g). Tensors that can be expressed as the outer product of a vector set are known as rank-one tensors. The number of vectors within the set is the order of this rank-one tensor; therefore, a rank-one tensor can have any number of modes. Rank-one tensors exhibit a single pattern association with each mode, enabling straightforward interpretation.

Most tensors are more complex than rank-one tensors. Nonetheless, by expressing them as the sum of rank-one tensors (Fig. 1h), interpretation becomes significantly easier, since they can be understood as the combination of these rank-one individual patterns. Even if we do not represent the original tensor exactly, if a small number of patterns can closely approximate the original tensor and capture essential information, we can still gain insights into the overall trends. This process of breaking down a complex tensor into the sum of a few patterns is known as tensor decomposition or tensor factorization.

## A step-by-step guide on tensor decomposition

### Structuring the data into a tensor format

Organizing a dataset into a tensor requires recognizing the structure defined by the experiment. In the example presented in Fig. 1c, it is natural to use a three-mode tensor with organ, drug, and time point modes. Tensor order can extend beyond three dimensions if, for instance, each organ, drug, and time combination was performed across multiple assays (e.g., measurement of many genes or proteins).

Measurements can only be separated into a distinct mode when the mode's labels relate to a common experimental entity across which the data can be grouped accordingly[5]. For example, should multiple technical replicates for each condition be grouped in a separate mode? No, because the "Sample 1" replicate of cells from liver does not signify the same replicate as "Sample 1" of cells from skin. We may average these replicates during reformatting if their variation is not of particular interest. However, if these samples represent a common set of patients—"Sample 1" is the same for all cell types indicating that they came from the same individual—this justifies the inclusion of a corresponding mode. As another example, in single-cell analysis, when combining runs from different backgrounds, whether the clusters with the same label should be aligned depends on whether each cluster label holds the same meaning across backgrounds. If the cluster labels are assigned randomly (e.g., in *k*-means), they are not equivalent between runs, and therefore cannot form a separate "Cluster" mode. However, if the clusters can be identified based on cell surface markers and "Cluster 1" consistently represents the same cell type, this cell type mode is justified.

In a tensor format, the items representing positions along a mode are treated as independent. Therefore, the order of items on a mode is inconsequential to the tensor decomposition. For instance, switching the positions of "3 hr" and "12 hr" on the time point mode in the tensor in Fig. 2b does not affect the results. For longitudinal measurements where sometimes the time points cannot be aligned perfectly, compromises may have to be made (More on longitudinal measurements in "Constraints on the factors"). One approach can be binning, where similar time points of different samples are grouped into one category. For instance, if one individual only has samples at "3 hr", while another only has "4 hr", a "3-4 hr" category may be created to align them. Sometimes, several positions in the tensor may be left empty to maintain the data's logical structure (see "Missing data and imputation" on decomposition with missing entries).

Tensor decomposition can benefit from appropriate data preprocessing, such as mean centering, normalization, and transformation. For a three-mode tensor, chords are an extension of columns in a matrix, whereas slices are all values associated with a specific position along one mode (Fig. 2a). To avoid larger values overshadowing the variation of interest, especially when the measurements are not in comparable ranges, one can normalize or standardize values chord-wise or slice-wise, which respectively correspond to one type of measurement across all samples, and all numbers aligned to one sample (Fig. 2a). The concept of chords and slices works with any mode, but the normalization is usually performed on the subject/sample mode. Tensor factorization also assumes a linear scale, meaning that doubling the number always equals twice the effect. Measurements that are not on a linear scale should therefore be transformed accordingly before the decomposition.

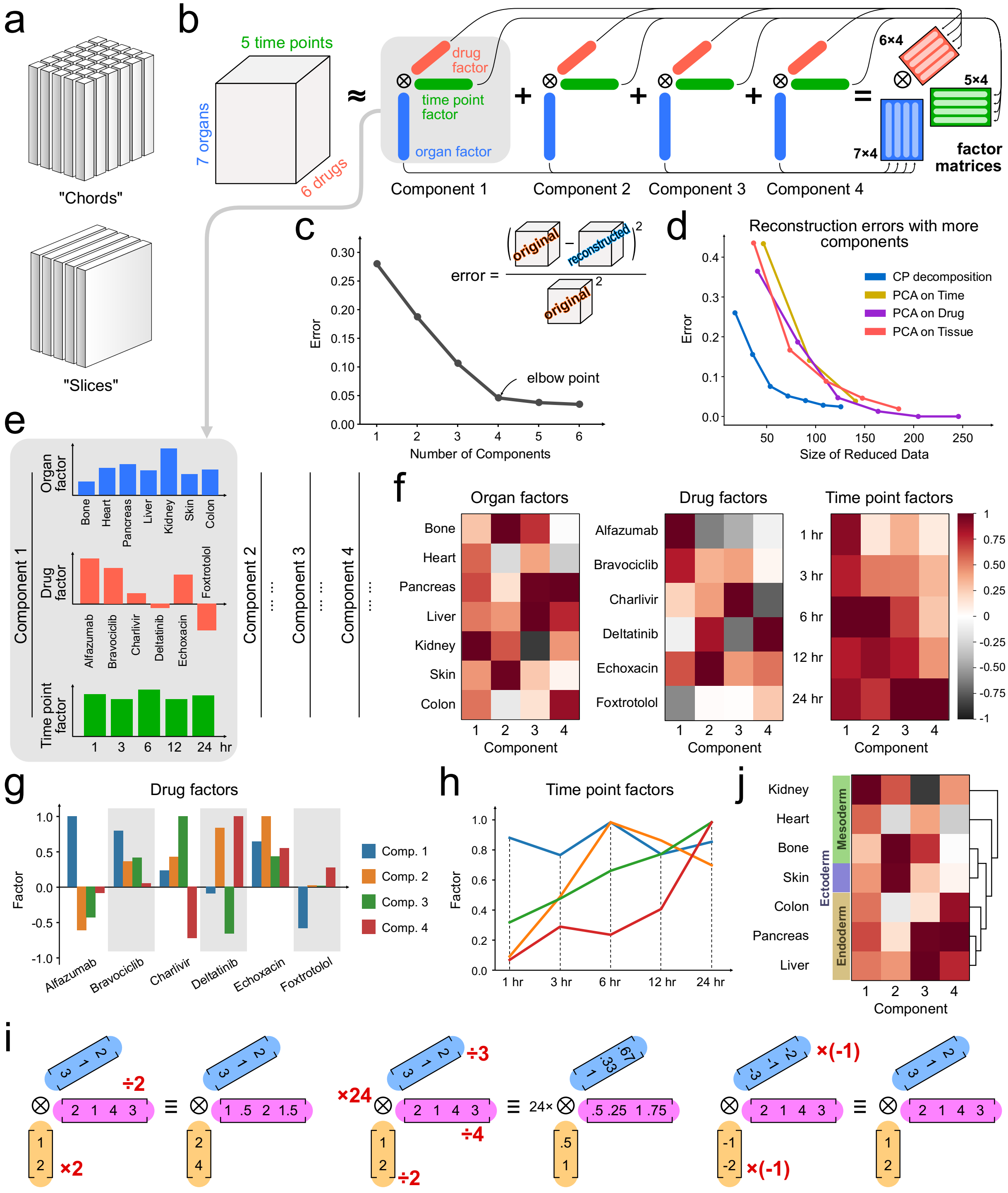

**Figure 2. Fundamentals of Canonical Polyadic (CP) tensor decomposition**

a) For a three-dimensional array, a chord is the entries across all labels on one single mode, and slices are entries across two modes. b) CP decomposition approximates a complicated tensor as the sum of a few rank-one tensors. In the example here, for a drug response tensor of 7×6×5, organ by drug by time point, after being decomposed into 4 components, we will have 4 factors for each of the three modes. Organizing them into matrices, we will have three factor matrices with shapes 7×4, 6×4, and 5×4 for organ, drug, and time point, respectively. c) Plotting the number of components against the error. Error is defined as the sum of squared differences normalized by the sum of squares of the original tensor. An optimal component number may be attained at the elbow point on the plot, or the point at which an acceptable error is reached. d) CP decomposition can represent the data more concisely than PCA when the data is structured appropriately for tensor. Sizes of reduced data are plotted against their reconstruction errors using CP or PCA unfolding the tensor in different directions. *(caption continues next page)*

e) Plotting every factor separately to visualize tensor decomposition results. Here, three bar plots demonstrate the three mode-specific factors of Component 1. The factors of other components are omitted but can be shown similarly. f) Heatmaps to visualize the factor matrices compactly. We can both interpret a component by inspecting all its factors across modes and/or comparing factors within a mode to distinguish their differences. g) Factors of a discrete variable mode (such as drug mode here) can be visualized with a bar plot. h) Factors of a continuous variable mode (such as time point mode here) can be visualized with a line plot. i) Demonstration of factors' scale indeterminacy. Scaling the factors coordinately (left), factoring the weights to a separate scalar (middle), or negating factors in pairs (right) all yield equivalent factorizations. j) Organ factor heatmap reordered by hierarchical clustering on the factorization results. Here, the reordered factors inadvertently align with the organs' biological grouping.

## Performing the decomposition

Here we introduce the decomposition method known as canonical polyadic (CP), parallel factors (PARAFAC), or canonical decomposition (CANDECOMP). Implementations of this method are available in software packages for various programming languages (Tbl. 1).

| Programming Language | Package | Decomposition methods | | Constraints Implemented |
|---|---|---|---|---|
| | | Introduced in this work | Other methods | |
| Python | TensorLy[6] | CP, Tucker, PARAFAC2, CMTF, CP partial least squares | Partial Tucker, Tensor Train, CP/Tucker regression | Nonnegativity, Symmetric, Regularization |
| MATLAB | Tensor Toolbox[7] | CP, Tucker | | Symmetric, Orthogonalized |
| R | rTensor[8] | CP, Tucker | 3-mode tensor SVD, multilinear PCA | |
| | Multiway[9] | CP, Tucker, PARAFAC2, core consistency diagnosis | Simultaneous Component Analysis | Nonnegativity |

**Table 1. Selected tensor decomposition packages and the methods they implement.** CP: Canonical Polyadic Decomposition (also called PARAFAC or CANDECOMP); CMTF: Coupled Matrix-Tensor Factorization; SVD: Singular Value Decomposition; PCA: Principal Component Analysis.

CP decomposition requires a data tensor and the desired number of components. The component number is the number of rank-one tensors used to approximate the original data (Fig. 2b). For each mode, the factors of each component can be grouped into a factor matrix (Fig. 2b, right), so that the first columns of each matrix represent the first factor, the second columns the second factor, and so on. Thus, if we take the outer product of the first columns (Factor 1) in the three factor matrices, we will obtain the first decomposed rank-one tensor, Component 1. Repeating this process for each component and summing them up, we can reconstruct a tensor that approximates the original data (Fig. 2b). To summarize, the decomposed factors can be either grouped by mode into factor matrices, or by the factor indices into components. The goal of the decomposition algorithm is to make this reconstructed tensor match the original one as closely as possible.

## The number of components

With CP decomposition, one must choose the number of components. Too few components will miss essential trends, while too many will lead to redundant factors, noise (overfitting), and poorer interpretability. One typically needs to experiment with a range of numbers to identify an optimal choice.

To quantify how well a decomposition with the chosen number of components fairly represents the original data, one can quantify the difference between the reconstructed tensor and the original data, or the fitting error. This value is calculated as the sum of squared differences between these two tensors, usually normalized by the sum of squares of the original data (Fig. 2c). Smaller errors indicate a better fit. While the error can range from 0 to any positive number, a successful fit should result in an error below 1 when normalized. The fitting error consistently decreases with a greater number of components, with diminishing returns where each additional component improves the fit to a lesser degree (Fig. 2c).

Achieving a perfect fit to the data is typically not the goal of tensor decomposition. While this is technically feasible by setting the component number equal to the tensor's theoretical rank[3], in practice, this number is almost always too high for any practical use. To choose the optimal number of components, one may identify where the benefit of adding more components diminishes. This sometimes corresponds to the kink (or elbow) point on the error plot. However, such a transition point is not always evident.

The process described above resembles selecting component numbers in PCA but with a few distinctions. Tensor decomposition is not a recursive process: the components of a 3-component decomposition are not necessarily a subset of the 4-component decomposition. Components are also not ordered. Therefore, to create an error plot, the decomposition must be run for each number of candidate components (Fig. 2c).

The choice of component number directly relates to the data compression efficiency and fidelity trade-off (Fig. 2d). Since a tensor can be approximated with its factorization results which consist of fewer numbers, tensor factorization effectively compresses it. The smaller the size of the reduced data, the better the data compression ratio. However, this comes with the cost of a worse approximation (i.e. a larger reconstruction error) of the original data. For example, for a tensor with $6 \times 7 \times 5 = 210$ values, a 4-component decomposition will compress it into $4 \times (6+7+5) = 72$ numbers; if using 3 components, only $3 \times (6+7+5) = 54$ numbers, but sacrificing the reconstruction error. When the data is structured appropriately for tensor analysis, CP decomposition can represent the data more concisely than PCA—achieving a smaller representation under the same fidelity, or comparable reduced sizes with lower error (Fig. 2d).

Another consideration is the inherent noise present in biological measurements. With too many components, tensor decomposition starts to fit trivial patterns which are more likely to be noise. In principle, we should cease adding more components when the algorithm begins to overfit (fit the original data too closely but lose generality), is prone to excessive local minima, or starts to violate the properties of CP. These situations may be assessed through imputation tests (see "Missing data and imputation"), factor similarity tests (see "Optimization algorithms"), or core consistency diagnostics (see "Tucker decomposition").

## Interpreting the results

After validating the decomposition, the resulting factors can be inspected for biological insight. To provide a concrete example, we read into our decomposition results shown in plots (Fig. 2e-h).

To visualize the results, one should design plots that describe how each factor is associated with the labels along each mode. Therefore, one can have one subplot for each factor (one from each mode) for each component, repeated for all components (Fig. 2e). In these plots, the x-axis indicates the labels, and the y-axis shows the factor weights. For a more concise visualization, one can also plot each factor matrix made from factors from all components as a heatmap with colors representing the weights (Fig. 2f). Generally, bar plots and point plots work well for discrete labels such as samples, cell lines, or molecules (Fig. 2g, left), while line plots are more suitable for continuous labels such as time or concentration (Fig. 2g, right). Overall, visualization should optimally serve the presentation of the insights, and there is no fixed rule.

The initial phase of interpretation involves delineating the meaning of each component pattern. This requires reading the plots across all modes. For instance, consider Component 1 (Fig. 2e). Within the organ factor, the largest signal originates from cells collected from kidney, followed by smaller weights from cells collected from the pancreas and heart. The same information can be also captured from the first column of the organ factor heatmap (Fig. 2f, left). Along the drug mode, the strongest signal appears on alfazumab, then bravociclib and echoxacin. This can be read out from the heatmap (Fig. 2f, middle) or the drug factor bar plot (Fig. 2g). The time point mode factors, on the other hand, all have relatively high values in Component 1. The blue line in the time factor plot best represents this trend (Fig. 2h). Putting this information together, one concludes that this component delineates a temporally persistent impact of alfazumab (and lesser so bravociclib and echoxacin) on cells from kidney (heart and pancreas somewhat as well). In practice, one can choose whichever plot best depicts the trend. Following the same logic, we see that Component 2 unveils an effect of mostly echoxacin on cells collected from bone and skin, peaking at 6 hours. Component 3 indicates an effect of charlivir on mostly cells from pancreas and liver that increases over time, while Component 4 also indicates an accumulating effect, but of deltatinib on cells from the pancreas and colon.

A specific mathematical intricacy, scale indeterminacy, can hinder clarity (Fig. 2i). As the effect along each mode is multiplied together, scaling these factors in an opposing way, i.e., doubling one factor and halving another, yields equivalent results (Fig. 2i, left). This indicates that only the relative ratios of weights within a factor are certain, not the absolute values. Therefore, we should not compare the absolute weights between factors of different components, only the relative composition. To avoid ambiguity, one typically normalizes all factors to a defined scale, storing the weighting as a separate scaler (Fig. 2i, middle). The issue of indeterminacy extends to negative factors. By the same logic, negating two factors simultaneously also yields equivalent results (Fig. 2i, right). This is sometimes called sign indeterminacy[10]. One approach to avoid ambiguity is to make most modes positive by negating the factor vectors in pairs, ensuring that at most only one mode harbors factors with the strongest negative values (Fig. 2i, right).

One can also compare across components within a single mode. Within the organ mode, for instance, Components 3 and 4 assign similar factors to cells from the pancreas and liver, unveiling shared localization in drug effect (Fig. 2f, left). In the drug mode, alfazumab and bravociclib have similar factors, suggesting that they have analogous interaction profiles (Fig. 2f, middle, 2g). The time mode also separates into three trends, ranging from stable (Component 1) to increasing over time (Components 3 and 4) and peaking (Component 2) (Fig. 2h). To better identify similar entries (e.g. drugs or organs) on a mode discovered by tensor decomposition, one can also perform hierarchical clustering on the factor matrix and reorder the entries accordingly (Fig. 2j). This juxtaposes entries of similar factor weights, helping to reveal groupings of comparable entries. Clusters may reflect biologically significant grouping (Fig. 2j).

## Details and considerations of tensor decomposition

The previous section presented an overview of employing tensor decomposition. However, several details of the procedure may help in certain circumstances.

### Optimization algorithms

Solving tensor decomposition is, in its essence, an optimization problem. The objective is to find a set of factor matrices that, when multiplied, render a reconstructed tensor with minimal error (Fig. 3a). Common mathematical optimization algorithms, such as gradient descent or the Newton-Raphson method, can be employed here[11]. This “direct optimization” approach offers the advantage of versatility, since many optimization methods allow additional constraints, making it possible to develop new decomposition schemes. However, its performance relies heavily on the chosen method and initialization values, since a substantial number of parameters must be simultaneously solved.

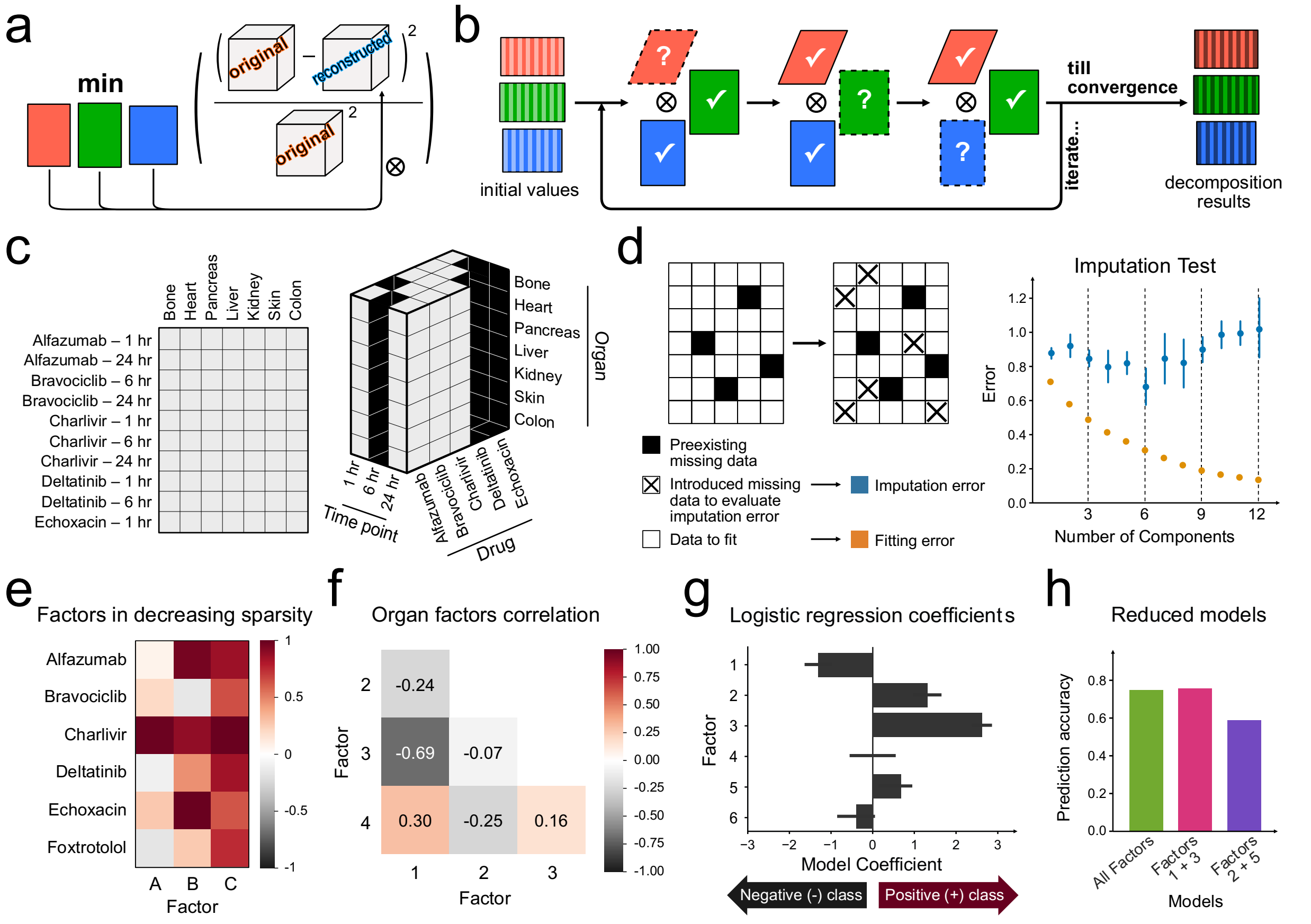

**Figure 3. Technical details on applying CP to biological data**

a) Solving tensor decomposition is an optimization problem aiming to minimize the reconstruction error by adjusting the numbers in the factor matrices. b) Alternating least squares (ALS) is another strategy besides direct optimization. Starting from a set of initial values, it optimizes one factor matrix at a time with linear least squares while holding the others constant. This process is repeated on each factor matrix until convergence is reached. c) A demonstration of how structuring data into tensor format may create missing values. Although the original table on the left does not contain any missing values, since not all drug-time pairs are measured, the reformatted three-mode tensor contains missing chords. d) Demonstration of the imputation test. Ignoring the preexisting missing data, we arbitrarily introduce more missing positions, use the remaining data to fit the decomposition, and then compare the reconstructed (i.e. imputed) values with the original values in the positions we removed. Plotting against the number of components, the fitting errors should decrease monotonically with more components, but imputation errors do not necessarily. e) Sparsity in factors. These fabricated Factors A, B, and C are in the order of decreasing sparsity. f) The correlation matrix of organ factors reveals their associations. g) Tensor decomposition factors can be used for response prediction when combined with regression. The coefficient of each factor indicates their association with the sample classes. h) For classification, the model can be reduced to using a subset of the factors.

As an alternative approach, we can first notice that the factor matrices exhibit symmetry: swapping mode orders does not change the solving. Also, if we know the correct factors of all other modes, solving for one mode can be converted into an ordinary least squares problem. Thus, we can tackle one mode at a time using least squares while treating the others as constant, then repeat this for every mode (Fig. 3b). We keep iterating until these factors converge. Over time, we can expect a monotonic decrease in the fitting error. This approach is called alternating least squares (ALS). Besides its efficiency, ALS often benefits from more stable and reproducible performance[12].

Both methods require initial factor values. While a random initialization may be sufficient, a more informed estimation can expedite convergence. One such estimation involves using the principal components from a flattened version of the original tensor. This approach, known as SVD (singular value decomposition) initialization, usually yields more stable results and reduces the likelihood of a suboptimal solution (i.e. local minimum). However, neither initialization guarantees the best solution.

When the resulting factors are highly dependent on the starting point of the fitting, it can indicate that the optimization problem is ill-formed, suggesting that the chosen number of components is too large or that additional constraints would be helpful. This property is exploited by the factor similarity test to determine the appropriate component number[13,14]. In essence, this test quantifies to what extent different starting points change the resulting factors, helping determine up to how many components the factorization algorithm remains stable.

## Missing data and imputation

Missing measurements frequently arise from experimental limitations. These omissions are not necessarily a result of oversight; certain measurements may be intentionally missing. This issue becomes particularly pronounced with tensors, as complete tensors require all possible combinations of all modes. Consequently, missing data can emerge simply from transforming a dataset into a tensor, even if the original data appears complete (Fig. 3c). For instance, the example dataset in Fig. 3c does not contain any missing values but, because the impacts of deltatinib and echoxacin after 6 hours were not measured, the reformatted tensor contains missing chords (Fig. 3c, right).

Tensor decomposition provides an avenue to impute the missing values of a tensor. Since a full tensor can be reconstructed from the resulting decomposed factors (Fig. 1g, 2b), one can use these reconstructed values from tensor decomposition to replace the missing positions, effectively imputing them[15]. Compared with matrices, higher-order tensors benefit from the additional information from more shared coordinates. To perform decomposition on an incomplete tensor, we simply disregard the missing positions and only fit the existing ones: in direct optimization, the optimization function ignores the missing positions; in ALS, some form of value in-filling or censoring is employed. Note that zeros in a tensor will still be fit by the tensor decomposition algorithm, unlike explicitly missing values, so replacing missing values with zeros is incorrect.

Tensor imputation through decomposition is not foolproof; it remains an area of ongoing research. Like matrix completion, it relies on inherent assumptions. If the original data cannot be approximated as the sum of a few rank-one tensors (Fig. 2a), the imputed values can significantly deviate from their true values. Other factors, such as the quantity and distribution of the missing values and the chosen component numbers and decomposition method, can also influence the accuracy of imputation. A tensor cannot be missing all its values across a slice. Thus, in situations where there are very few non-missing values, it may be advantageous to consider discarding one position along a mode.

One can use imputative performance to assess the reliability of tensor decomposition or to determine the appropriate decomposition rank. In an imputation test, one intentionally introduces randomly posited additional missing values in the data (Fig. 3d, left). Following decomposition, the entire tensor is reconstructed from the factors, and the left-out values are compared against their reconstructed versions. A substantial disparity indicated by a high imputation error indicates an unsuccessful decomposition, attributable to either an ill-suited decomposition or an excessively high number of components. While the fitting error monotonically decreases with more components, the imputation error often shows an optimum at an intermediate number of components (Fig. 3d, right).

## Constraints on the factors

The optimization processes introduced so far only aim to fit the data. However, their results may suffer from low interpretability, overfitting, and instability. Numerical constraints on the factors can help with these issues. Although they may impact the goodness of fit, reasonable constraints can enhance the model's ability to reveal meaning patterns, leading to more insightful discoveries. For example, one goal of constraints is to achieve sparsity, where a factor has nonzero values in a few positions and renders others nearly or exactly zeros. This helps establish direct associations between factors and their effects[16]. For instance, in the hypothetical drug factors in Fig. 3e, sparse Factor A uniquely associates with charlivir and isolates its effect from other drugs. The less sparse Factor B is relevant to at least three drugs, while Factor C has similar values across all drugs, unable to separate the effect of one from another.

Nonnegativity is the most commonly used tensor decomposition constraint[17]. It aligns intuitively with the expectation that certain quantities in biology are inherently nonnegative: a cell cannot secrete a negative number of molecules, and a gene cannot be expressed at a negative level. Nonnegativity also serves to foster sparsity within the factors and avoid overfitting. Decompositions allowing negative factor values can yield degenerate components, where one component is strongly positive and another is strongly negative, mostly canceling each other out[18]. Enforcing all values in the factor matrices to be nonnegative obviates such occurrences, as the impact of any component cannot be counteracted by another. Nonnegative factorization often leads to minimal increases in model error, solidifying its application in practice[13].

Constraints can also be used to enforce biological knowledge in a decomposition[19]. For instance, in neuroscience, one may postulate minimal crosstalk among different brain regions and limit the brain region factors to be a diagonal matrix[20]. In molecular biology, one may employ orthogonalization of the factors to enforce a clean delineation between components and traits[21]. This usually lacks a standardized approach, as biological contexts vary, and may require customized solving[14].

## Subsequent analysis

While tensor analysis often serves as an important step for distilling data into significant patterns, further analysis beyond plots (Fig. 2e-g) is often required to learn what component patterns indicate about biology. The factor matrices serve as efficient summaries for individual patterns linked to their respective modes. Consequently, each matrix can be isolated for a detailed analysis of the variation within a specific mode of interest. For instance, to better quantify the similarities between components, one may calculate correlation plots and matrices (Fig. 3f). For example, it may not be immediately apparent that Components 1 and 4 of the organ factors are positively correlated. Also, those components associated with genes or molecules of particular interest from prior knowledge can be further examined to validate their agreement with known mechanisms.

The decomposed factors can also be used as reduced data to predict responses or sample classes when combined with regression. The scale and sign of the weights for each factor indicate its effect on the regressed quantity (Fig. 3g). If not all factors contribute to the effect of interest, the prediction model may use only a subset of them (Fig. 3h). For example, in prediction using only two factors, Factors 1 and 3, performs just as effectively as all factors (Fig. 3h).

## Advanced tensor methods beyond CP

In this section, we delve into more advanced tensor decomposition methods. For more complex biological data, it is particularly crucial to choose a method that best reflects the structure of the expected patterns.

### Tucker decomposition: allowing all factors to interact

In CP decomposition, especially when there are more components, some components may start to look similar along one mode. For example, in the previous example, the Components 3 and 4 time point factors appear similar, both exhibiting an increasing trend over time (Fig. 2f, right). This redundancy arises from the inherent constraint of CP decomposition, where factors may only interact the same component (Fig. 2b). In other words, because CP does not allow interaction between Component 4 of the drug factor and Component 3 of the time point factors, a repetitive time factor must be present in Component 4 to capture a similar rising effect over time. CP permits the existence of two identical factors in one mode, as long as their corresponding factors in other modes remain distinct. This assumption compromises data compression efficiency. Therefore, it can be beneficial to relax the rigid structure of CP decomposition.

In Tucker decomposition, all factors across modes are allowed to interact[22,23]. For example, here we perform Tucker decomposition (Fig. 4a) on the same 7×6×5 drug response data tensor shown in Fig. 2b. In contrast to CP, Tucker decomposition permits varying numbers of factors for each mode. In the present scenario, we recognize 4 distinct organ factors, 2 drug factors, and 3 time factors. This flexibility allows for interactions between all factors across modes, eliminating the concept of a distinct "Component 1." Consequently, it becomes necessary to scrutinize all 4×2×3=24 potential interactions among the factors in the three modes. The magnitude of each interaction is characterized by its corresponding weight, with zero indicating the absence of interaction. These 24 weights can be arranged as a tensor known as the core tensor with dimensions of 4×2×3 (Fig. 4b). For instance, the core tensor in Fig. 4b reveals a substantial interaction among organ Factor 1, drug Factor 1, and time Factor 1, while organ Factor 2, drug Factor 2, and time Factor 1 exhibit minimal interaction. Therefore, this factorization is denoted as a (4,2,3)-rank Tucker decomposition. The outcomes of a Tucker decomposition include a core tensor and three factor matrices corresponding to the three modes. Each factor matrix assumes the shape determined by the number of entries on the mode in the data and the number of factors associated with that mode (Fig. 4a, bottom).

CP decomposition is equivalent to a specific instance of the Tucker decomposition, wherein factors associated with different components are non-interacting, thus the core tensor assumes a superdiagonal form, signifying that all off-diagonal positions are zeros (Fig. 4c). This superdiagonal property has been harnessed to test whether a CP decomposition is correctly implemented, known as core consistency diagnosis[24]. Specifically, after acquiring the CP factor matrices, if adding off-superdiagonal interactions to the core tensor can improve the fitting considerably, the number of components may be inappropriate, or Tucker is a better model than CP for this dataset.

The flexibility provided by Tucker decomposition offers a concise representation of biological data. However, with increased flexibility comes challenges in interpretation. Unlike CP, where each component cleanly represents a single pattern, understanding Tucker's results requires examining all potential factor interactions.

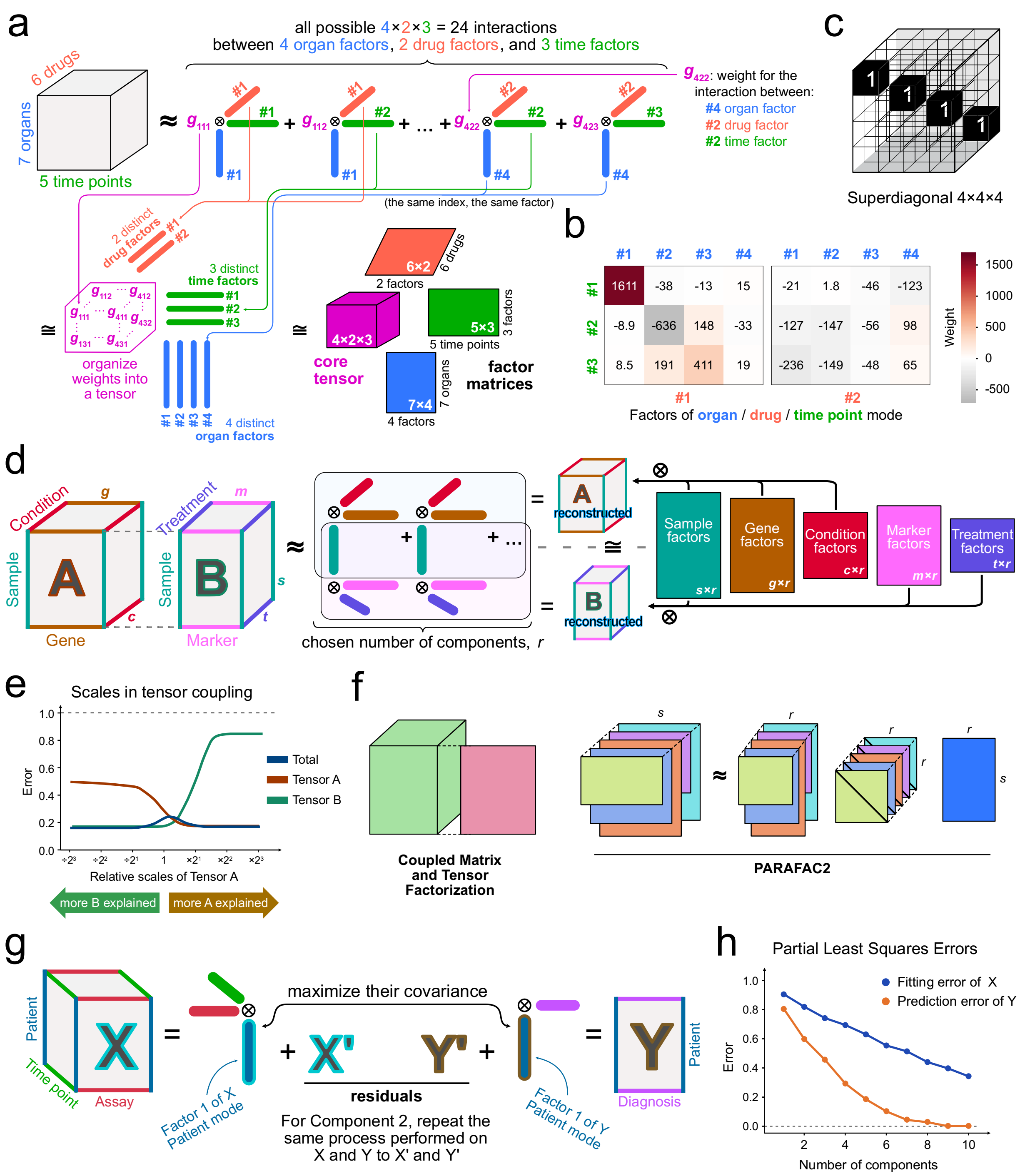

**Figure 4. Tensor methods beyond CP: Tucker decomposition, coupling, and partial least squares**

a) Schematic of Tucker decomposition. This (4,2,3)-rank Tucker decomposition on the previous 7×6×5 drug response tensor allows all distinct 4 organ factors, 2 drug factors, and 3 time factors to interact. The weights of these 24 interactions are organized into a 4×2×3 core tensor. The results of a Tucker decomposition are a factor matrix for each mode and this core tensor. b) The core tensor of a Tucker decomposition. It can be visualized by showing the numbers in each slice. c) A superdiagonal 4×4×4 tensor. CP is a special case of Tucker decomposition where the core tensor is superdiagonal. d) Schematics of coupled tensor decomposition. Here, two three-mode tensors, A and B, are coupled on the sample mode. Therefore, the sample factors are shared, while the gene and condition factors are private to Tensor A, and marker and treatment factors Tensor B. The dimensionalities of them are indicated by the lowercase letters. e) The scaling issue in coupled tensor decomposition. When one of the coupled tensors has values with greater total variance, the factorization explains more variance in it if without proper scaling, leading to an uneven representation of the two datasets. f) Some other examples of tensor coupling: coupled matrix and *(caption continues next page)* tensor factorization (CMTF, left) and PARAFAC2 (right). g) Schematics of tensor partial least squares. Partial least squares is performed on two tensors, X and Y, with one aligned mode. During solving, the two separated X and Y factors of the aligned mode (patient mode in the example case) yield the maximal correlations. Partial least squares components are solved sequentially, as the next component is found by repeating the same process on the residuals, X' and Y', from the last round. Therefore, the components are in decreasing order of covariance explained. h) The performance of partial least squares can be evaluated by calculating the fitting errors of X and the prediction errors of Y. Both should be decreasing with more components at diminishing rates till they reach zero.

## Coupling: sharing factors across multiple tensors

The integration of (epi-)genomic, transcriptomic, and proteomic data, either in bulk or at the single-cell resolution, has provided opportunities for an integrated understanding of cellular processes. In a broader sense, biologists often encounter data fusion challenges when attempting to identify shared patterns among multiple data sources[25]. The joint analysis of several datasets can be formulated as coupling of tensors[26].

Coupling arises when two or more datasets are collected with differing dimensions, but all tensors share at least one "coupled" mode (Fig. 4d, left). Commonly coupled modes include samples/patients or genes that are shared across multiple assays. For instance, there may be two datasets on the same group of samples, one of gene expression under various experimental conditions (Fig. 4d, Tensor A), while the other of cell surface markers over various treatments (Fig. 4d, Tensor B). In this case, the sample mode is shared, while each tensor has other uncoupled modes, such as genes and surface markers. In a coupled decomposition, a shared mode will have a common factor matrix that is used by all tensors that comprise this mode. In this way, this factor matrix succinctly reflects the trends across the coupled tensors.

Visualizing and interpreting the results of a coupled tensor factorization operate like with CP (Fig. 2f). Tensors are decomposed into a series of rank-one components, and any coupled mode will have a single set of factors shared among all the tensors using it (Fig, 4d, middle). Each distinct mode will still have a factor matrix (Fig, 4d, right). In addition to examining components within a tensor, one can also compare the uncoupled private modes between two tensors to assess their associations. A unique advantage of coupling arises from missing data. If a certain tensor has missing entries, other tensors can share information through the coupled factors to improve imputation.

Coupling introduces a new issue. Because factorization minimizes the overall error, the priority in explaining patterns from each dataset is influenced by their relative scaling. As the total variance of values can differ significantly across tensors of various assays, the decomposed factors can be dominated by one source if the data is not appropriately scaled. Typically, a range of scaling should be explored, and the overall and tensor-specific error evaluated (Fig. 4e). If the factor matrices will be used to predict some outcome of interest, the prediction accuracy using factorizations from various scalings can be explored.

Overall, coupling offers remarkable flexibility for all types of biological data. Although we refer to the methods as coupled tensor factorization, matrices (2-way tensors) are also included. For example, many applications have used coupled matrix and tensor factorization to jointly analyze a tensor and a matrix (Fig. 4f, left)[27,28]. Coupling also expands the applicability of tensor methods to more irregularly shaped data, as illustrated by PARAFAC2[29]. PARAFAC2 is a method that decomposes a series of matrices, where one mode is shared while another is unaligned and variable in size (Fig. 4f, right). This forms a ragged tensor to which CP or Tucker cannot be applied. PARAFAC2 projects the variable modes into a latent, uniform shared mode, identifying patterns not only on the shared mode but also across these matrices, effectively harnessing the benefits of coupling (Fig. 4f, right). Tensor coupling is an active field of method development, including combining it with other decomposition strategies (such as Tucker or partial least squares).

### Partial least squares: informing decomposition by effects

Many scientific questions involve identifying how a series of measurements associate with a specific phenotype or outcome of interest. For example, one might associate patients' blood panel tensor with their diagnosis. In statistical terms, we have explanatory variables (X) and outcomes (Y), and our goal is to reveal only the patterns in X that uniquely associate with Y. This approach differs from simply coupling them where the joint variance of both X and Y is considered. Instead, the objective is to only capture the trends in X when they exhibit correlation with Y.

As mentioned previously, tensor decomposition factors can be combined with linear regression models. This two-step approach bears a resemblance to principal component regression: first, the data is decomposed using tensor decomposition without considering the effects; then, regression is applied to capture correlations between the decomposed factors and their effects (Y). However, as the first step is performed without the knowledge of Y, the decomposed X factors are not guaranteed to associate with Y. To address these challenges, partial least squares (PLS) methods have been developed, in both classification form (PLS discriminant analysis) and regression form (PLS regression)[30].

Tensor PLS is designed to uncover relationships between two tensors, X and Y, for predictors and responses, wherein one mode is aligned (Fig. 4g). For instance, consider tensor X representing medical tests on a group of patients over time, while matrix Y (a two-way tensor) records their diagnosis. The result of tensor PLS is analogous to performing two separate CPs on both X and Y simultaneously with the same number of components. After decomposition, they will each have a distinct patient factor matrix. However, PLS decomposes both datasets with the goal of maximizing the correlations between these two patient factors (Fig. 4g). The factors of the other non-aligned modes in X and Y come after obtaining the patient factors, and are defined to maximally capture variance within each dataset[31]. While the intricacies of the solving algorithm extend beyond the scope of this introduction, one helpful property to note is that tensor PLS is solved component-by-component. Each additional component is solved upon the residuals of X and Y (X' and Y') which are the original tensors subtracted by the solved components (Fig. 4g), meaning that components are ordered by the covariance they explain. To assess the fitting performance, one can calculate errors for both X and Y, and these errors should be monotonically decreasing when adding more components (Fig. 4h).

Overall, PLS has unique advantages when focused on a particular response. Since it is designed to specifically discover those patterns associated with a prediction of interest, PLS can predict the effect with fewer components compared with CP. Tensor PLS can be combined with Tucker decomposition and coupling in explanatory (X) tensors, and specific techniques are available to handle missing values[32].

## Biological insights from tensor-based methods

Tensor decompositions have applications in virtually all fields of biological data analysis. In this section, we delve into several exemplary works that effectively showcase this.

### Applications in bioinformatics

In bioinformatic studies, multi-omics data may contain tens of thousands of genes and millions of genomic positions. Tensor methods can simplify these large datasets generated by high-throughput techniques into a succinct set of components and do so more efficiently than matrix-based counterparts.

These reduced latent structures group genes based on their common patterns revealed by the data, easing the scale of effect prediction.

Hore et al. illustrated how tensor methods can be applied to condense genes in RNA-seq data across multiple tissues into associated factors to reduce the scale of statistical testing and to strengthen their statistical power[16]. To reveal gene networks, they structured the gene expression levels into a gene by individual by tissue tensor. After applying the tensor method, the data was reduced into around two hundred components, a great reduction from the tens of thousands of genes they originally dealt with. These components grouped the genes by activities and indicated in what tissues they were active. Using individual scores as genotypes for genome-wide scanning on SNPs, they discovered the components that were significantly associated with *trans*-expression quantitative trait loci (eQTLs) and revealed their specific pathway or epigenomic regulation.

Using tensor factors to cluster genes in transcriptome is further exemplified by Wang et al[10]. With the increasing scale of multi-tissue datasets, classical clustering methods struggle to extract information from multi-way interactions in the transcriptome. To fully extract the three-way interactions between individuals, genes, and tissues, they applied constrained CP to RNA-seq and microarray measurements. Besides being able to run on three-dimensional data where traditional methods failed to reveal true patterns in simulated data, this tensor-based clustering method was shown to better test for differentially expressed genes with improved statistical power compared with single-tissue tests.

Durham et al.[33], on the other hand, applied tensor methods to large epigenome projects such as Encyclopedia of DNA Elements (ENCODE)[34] and the Roadmap Epigenomics Project[35]. In these massive datasets, many cell type and assay pairs were not measured due to time and funding constraints. Therefore, the imputation of these data has been extensively studied[36]. Organizing the ENCODE data into a three-mode tensor, they found that tensor-based imputation outperformed alternative approaches, demonstrating that structuring the data in tensor form helps model and explain variation across the data.

Other tensor methods have been applied to epigenomic data too. For example, a variant of Tucker decomposition has been applied to model spatial association within topologically associating domains[37]. The decomposed factors directly link epigenomic state and chromosomal topology. Tensor decomposition can be also combined with machine learning methods. For example, extending the work of Durham et al., the same group inputted the concatenated tensor factors from three different genomic resolutions into a feed-forward deep neural network to predict the epigenomic signals, allowing a multi-scale view of the genome[38].

## Applications in neuroscience

Neuroscience is among the earliest fields to employ tensor methods[39]. As electroencephalography and functional magnetic resonance imaging data are collected over time, any experiment involving more than one electrode and trial is guaranteed to be at least three-dimensional. Conventionally, the data has been converted into matrices by averaging multiple trials, inevitably losing information about trial-to-trial variation. Therefore, tensor methods, including both CP and Tucker decomposition, have been attractive to the neural signal processing community[40].

Williams et al. presented a clean framework for applying tensor component analysis on large-scale neural data across time and trials[13]. Before running on the actual data, they demonstrated that tensor decomposition works well on simulated linear model neural networks and nonlinear recurrent neural networks, separating positive and negative cells with almost perfect accuracy. With the same simulations, PCA and independent component analysis failed to recover the right signal. They then applied the method to their experiments on mice's prefrontal activity and primate motor cortex. Nonnegative tensor decomposition was shown to cleanly separate neurons that were activated in various periods and associated with specific movements.

### Applications in systems biology

Systems biology makes repeated measurements over different times, tissues, or spatial structures, so the data are naturally in tensor structure. These measurements may include sequencing, flow cytometry, or quantitative cell imaging, requiring solutions for data integration. Two specific concerns here are avoiding overfitting, as the datasets are often limited in size, and incorporating heterogeneous information. Therefore, nonnegative decomposition, imputation tests, coupling, and partial least squares have been used.

Tensor methods offers unique advantages for the study of systems biology by enabling concurrent comparison of multiple contexts and extracting their shared trends. For instance, Armingol et al. employed tensor decomposition to study cell-to-cell communication from RNA-seq data[41]. Contrary to many previous studies that cannot handle more than two cellular contexts simultaneously, by embedding communication matrices[42] into a four-mode tensor, they were able to characterize variation in cell-to-cell communication across several contexts coordinately.

The benefit of tensor decomposition in analyzing repeated measurements simultaneously can also be extended to compositional data in microbiology. Microbiome studies often take multiple samples from the same individual either longitudinally or spatially, but there is a lack of methods to account for both biological change and interindividual variability in them. Martino et al. took the tensor approach to deconvolute gut microbial sequencing data[43]. They demonstrated that unsupervised tensor decomposition can identify differentially abundant microbes, accounting for the high-dimensional, sparse, and compositional nature of microbiome data.

Tensor partial least squares can also be helpful in systems biology. Netterfield et al. recently applied it in a study of DNA damage response[44]. They systematically profiled a human cell line with the treatment of DNA double-strand break-inducing drugs over time and concentrations, using tensor PLS to directly associate signaling to response, both as three-mode tensors, separating the time mode from drug concentrations. This allowed them to identify signals with time-dependent correlations with senescence and apoptosis. They also observed that tensor PLS required fewer parameters to predict the response than the conventional unfolded version.

## Conclusion

In this work, we introduce the application of tensor decomposition to biological data analysis. The paramount lesson of this work is the profound influence of the chosen data representation, the “medium,” on our comprehension of the data itself and the analytical approach. The selection of data representation should be driven by the natural structure of the underlying data and experiment rather than mere mathematical expediency. Approaching this analysis appropriately improves on the insights one can derive from the data through better accuracy, more evident interpretation, and an enhanced ability to integrate data across studies and scales. While tensor methods have gained increased prominence, work in multidimensional data analysis is far from its culmination[45]. Part of the field’s maturation will arise from a broader appreciation and understanding of these techniques.

Nevertheless, tensor decomposition, in its current form, is not without limitations. First, it is still fundamentally linear, so it may fail on datasets of nonlinearity characteristics. This does not forbid it from being an adequate baseline model though. Furthermore, the existing solving algorithms continue to grapple with numerical issues such as nonuniqueness in factors, instability when addressing missing data values, and challenges in hyperparameter tuning. These issues will be resolved by new theories and a broader appreciation of these techniques.